\documentclass[review]{elsarticle}

\usepackage[T2A]{fontenc}
\usepackage[utf8]{inputenc}
\usepackage{amsmath}

\usepackage[english]{babel}
\usepackage{hyperref}
\usepackage{gensymb}
\usepackage{mathtools}
\usepackage{amsmath}
\usepackage{enumitem}
\usepackage{graphicx}
\usepackage{systeme}
\usepackage{multicol}
\usepackage{multirow}
\usepackage{longtable}
\usepackage{xcolor}

\journal{Acta Astronautica}

\begin{document}
\begin{frontmatter}
\title{Analysis of Orbital Configurations for Millimetron Space Observatory}

\author[1]{A. G. Rudnitskiy\corref{corref1}}
    \ead{arud@asc.rssi.ru}
\author[2]{P. V. Mzhelskiy}
\author[3]{M. A. Shchurov}
\author[4]{T. A. Syachina}
\author[5]{P. R. Zapevalin}
\address{Astro Space Center, Lebedev Physical Institute, Russian Academy of Sciences, Profsoyuznaya str. 84/32, Moscow, 117997, Russia}
\cortext[corref1]{Corresponding author}

\begin{abstract}
In this contribution a primary feasibility study of different orbital configurations for Millimetron space observatory is presented. Priority factors and limitations were considered by which it is possible to assess the capabilities of a particular orbit. It included technical and scientific capabilities of each orbit regarding the fuel costs, satellite observability, the quality of very long baseline interferometric (VLBI) imaging observations and source visibilities.
\end{abstract}

\end{frontmatter}


\section{Introduction}
\label{sect:intro} 
Millimetron space observatory is the next generation space mission that is currently under development and planned to be launched after 2029. The active phase of scientific program preparation for Millimetron space observatory started in 2019. This mission is aimed to study the Universe at different scales, starting with protoplanetary disks in the Galaxy to the study of subtle cosmological effects \cite{Kardashev2014}. To cover all the scientific tasks, the observatory will have two observing modes: the single-dish mode and the space-ground interferometer (space-ground VLBI). 

Scientific tasks for single-dish mostly correspond to the sub-mm and far-infrared (FIR) wavelength ranges. The tasks are divided into a set of directions: cosmology (observations of CMB spectral distortions), compact heavily obscured galaxy nuclei, study of filaments and magnetic fields on various scales, star and planet formation, gravitational lensing, Sunyaev-Zel'dovich (SZ) effect, Solar system and search for water trail and complex molecules in the spectra of various cosmic sources and origin of the life in the Universe. Single-dish mode will include a set of instruments aimed covering the wavelength range from 0.07 to 3 mm. Practically all scientific observations in the single dish can be done only in outer space, on board of a space telescope as the Earth's atmosphere imposes strong restrictions on the coherent integration time and introduces significant distortions in the signal especially. Thus, going to space for sub-mm and FIR observations is the only option. 

One of the main scientific goals of the space-ground VLBI mode is to observe the shadow and photon rings of black holes. One of the basic techniques in this case is to obtain images of black hole shadows, namely Sgr~A* and M87. However, the detection of intrinsic structure of Sgr~A* at extremely long baselines may be strongly limited by the scattering due to the observable substructure \cite{Gwinn2014,Johnson2016,Johnson2018}. Thus, going to the baseline projections larger than $2-3\times10^{4}$ km the substructure of the scattering disk of Sgr~A* will dominate the intrinsic structure of the source itself at 230 GHz \cite{Fish2020}. Another issue is related to the observations of Sgr~A* is its high time variability. According to the simulations presented in \cite{Moscibrodzka2014}, \cite{Moscibrodzka2016} the characteristic time variability is about $\sim 221$ s, which 1000 times less than for M87. Thus, resolving its dynamical structure requires good snap-shot $(u,v)$ coverage \cite{Palumbo2019,Raymond2021}. These facts apply additional restrictions to space-VLBI observations of the Galactic center.

High resolution imaging of SMBH with using space-VLBI is currently a very promising area of scientific research (see, for example, \cite{Fish2020,Hong2014,Roelofs2019,Andrianov2020,Petretti2021}), but they all consider only different near-Earth orbits for space telescopes. Imaging the black hole shadow with VLBI straightly depends on the angular resolution, $(u,v)$ coverage and sensitivity of telescopes. Recent experiments of the Event Horizon Telescope allowed to obtain the image of supermassive black hole located in M87 galaxy \cite{EHT2019_P6}. Further detailed study of black hole shadows and precise estimation of its parameters requires higher angular resolution \cite{Martin2019,Kudriashov2019}. For ground-based VLBI the resolution is straightforward limited by the size of the Earth. Determination of the black hole metric and their parameters is also directly related to the spatial resolution of the interferometer. The search task is complicated by the fact that observations must be carried out for a specific sample of sources. In this case the launch of a space radio telescope capable of operating at frequencies compatible with ground observatories is a straightforward way to significantly improve the spatial resolution.

The success of any space mission directly depends on its planning, in which the choice of the most optimal and suitable orbit plays a primary role. The choice of such an orbit for the space observatory should be based on the results of analysis of a set of parameters, which include both the technical aspects of the feasibility of a particular orbit configuration and the scientific potential. Such analysis includes the estimation of fuel costs, spacecraft observability, source visibility, baseline projections (i.e. angular resolution) at which the studied sources can be observed and $(u,v)$ coverage quality regarding the space-ground VLBI mode.

In this article, we would like to dwell in more detail on the VLBI mode of Millimetron space observatory as one of the main issues of VLBI and, in particular, space-VLBI is the geometry of interferometer that is formed by the telescopes located on the Earth and in space. Obviously, the geometry of space-ground interferometer will depend on the orbital configuration of the space observatory and the total number of participating antennas. For these purposes, we considered and calculated several orbital configurations motivated by the experience of previous missions, technical and scientific requirements. The detailed description of these requirements, the motivation and the selected orbital configurations is provided in Section \ref{section:orbits} and Section~\ref{section:orbits_calc}.

\subsection{Millimetron Space Observatory}\label{subsection:observatory}
Millimetron space observatory will have 10 meter deployable antenna capable of observing at far-infrared, sub-millimeter and millimeter wavelengths. In order to achieve the best sensitivity for bolometric detectors of the single-dish mode instruments, the antenna will have active and passive cooling systems with cooling down to 10~K and the scientific payload will be located in the container actively cooled down to the temperature of 4~K.

Single-dish mode will have three instruments and together they will cover the frequency range from 100 GHz up to 6.0 THz with the sensitivity of detectors down to $10^{-17}-10^{-22}$ W/$\sqrt{Hz}$.

VLBI is going to cover the frequency bands from 86 to 320 GHz with additional bands of 43 and 570 GHz to be confirmed. The selection of such frequency ranges for Millimetron space observatory is based on the compatibility with ground observatories and proposed scientific cases. The need in VLBI observations at higher frequencies is dictated by the fact that the close surroundings of black holes should be optically thin in order to be imaged. Additionally, this will minimize the contribution from scattering effects to the resulting images of the Sgr A* black hole shadow in particular \cite{Gwinn2014,Johnson2015,Johnson2016,Johnson2018}. Such high frequencies will be available due to the hydrogen atomic clock with an Alan deviation of about $\sigma(\tau) = 1.9\cdot10^{-15}$ on $\tau\sim100$ s that will be located on board of the observatory \cite{Demidov2018}.
The receivers are expected to be mounted together with a quasi-optical system same as represented in Korean VLBI Network (KVN). Such system will provide the capability to perform simultaneous multi-band VLBI observations \cite{Han2012,Han2017}.

Currently, the design sensitivity of VLBI frequency ranges is estimated to be 1100, 2700, 5500 and 7200 Jy for 43, 86, 120, 240 and 345 GHz correspondingly. The expected detection limit of 5$\sigma$ on the baseline Millimetron-ALMA is 5 and 35 mJy for 10 and 1 s of integration time correspondingly for total bandwidth $\Delta\nu=4$ GHz. The final sensitivity will be known after testing of the on-board receivers performance and antenna efficiency.

More information about the instrumentation, technical parameters and requirements of Millimetron space observatory can be found on the official web-site\footnote{\url{https://millimetron.ru/en/}}.

\section{Constraints and Choice of Orbits}\label{section:orbits}
The choice of orbital configurations must be based on technical and scientific requirements and constraints of the mission.  First of all, one should take into account the planned lifetime of the observatory. This determines the necessary fuel costs for the formation and maintenance of the mission orbit. For Millimetron the total lifetime is 10 years with the guaranteed cryo operations of 3 years. 

As the antenna and the scientific payload of Millimetron space observatory will be mechanically cooled down to 10~K and 4~K correspondingly, it is important to avoid direct exposure of the antenna to the Sun, to Moon and the Earth. This leads to the constraints of the spacecraft orientation and thermal regime. The requirements for viewing angles "Sun-Spacecraft-Source" of Millimetron space observatory are not less than 95$^{\circ}$. At the same time the orbit should be selected such as to exclude the shadows from the Earth and the Moon, because of the limited capacity of the onboard power supply systems.

Additional constraints on the orbital configuration and its period will arise from the maximum time of continuous observation, the data transfer rate from the onboard memory and the volume of the memory itself. These limitations are primarily critical for the VLBI mode and they must be taken into account when planning a mission. We considered a 100 Tb onboard memory volume which correspond to $\sim$ 1700 minutes of continuous VLBI observations.

The current implementation of onboard data downlink channel will be capable to transmit the information from the onboard memory storage with a rate of 1.2 GBit/s. It means that it will take about 180 hours of continuous data transmission to download 100 Tb of the data from the space observatory. This is a critical point that is now being considered for the upgrade and corresponding studies are being performed for it. However, this brings additional orbital requirements to maximize the spacecraft visibility from the ground stations in order to extend the possible data transmission time.

Moving to the the science cases, the orbit should have a periodicity that would allow repeated observations of the same areas of the sky both in the single-dish mode and in VLBI mode. In addition, the orbit should cover as much of the celestial sphere as possible during the year. For the VLBI mode, the orbit should form a suitable $(u,v)$ coverage that would would satisfy the scientific tasks (imaging Sgr~A* and M87).

Summarizing the above constraints and requirements, we can conclude that the observatory must necessarily operate in orbit around the L2 point of the Sun-Earth system since the onboard active cooling system of the Millimetron space observatory can provide the required thermal regimes of antenna and scientific payload for the cryo period of the observatory operation (3 years) only there. 
At the same time, we considered additional near-Earth orbits (primarily to improve the VLBI observations) as the second stage of the mission under the assumption of returning the spacecraft from the L2 point after completing the scientific program requiring the best sensitivity and deep cooling of the observatory.

It would be an oversight before choosing specific orbital configurations not to mention previous space missions that were operating around L2 point and on the near-Earth orbits (presumably in VLBI mode). We took the experience of HALCA, Astro-G and Radioastron the near-Earth orbits. HACLA and Astro-G orbits could provide $(u,v)$ coverage of satisfactory quality for imaging observations \cite{Hirabayashi2000,Tsuboi2009}. High elliptical orbit that was used in Radioastron mission had an apogee of $\approx$ 330,000 km, which made it possible to achieve the highest angular resolution of 8 $\mu$as at the shortest wavelength of the mission ($\lambda=1.3$ cm) \cite{Kardashev2014_2,Zakhvatkin2014}. Earlier scientific space-VLBI missions such as HALCA (VSOP) and Radioastron have shown the principal possibility of conducting interferometric observations with ground-space interferometers. Recent studies demonstrated the capabilities and advantages of the middle near-Earth circular orbits, namely for space-space VLBI with two or more space radio telescopes (see for example, Event Horizon Imager concept \cite{Roelofs2019}). However, before the Millimetron space observatory, the question of the applicability of the VLBI technology to spacecraft in orbits located around the Earth-Sun L2 point has never been raised. This article discusses the very possibility of conducting VLBI observations in such (L2 point) an orbit.

As for the orbital configurations in the vicinity of the Lagrange point L2 of the Sun-Earth system, the following missions can be noted as example of successfully operated and operating space observatories at L2 point: WMAP, Planck, Herschel, GAIA and the upcoming James Webb Space Telescope, WFIRST and others \cite{WMAP2003,JWST2006,Planck2010,Herschel2010,Gaia2016,Wfirst2018}. As we mentioned about the possibility of a two stage mission, a special attention should be paid to the lunar mission CHANG'E-2 which with the help of a maneuver successfully reached point L2 in 2011 and became the first spacecraft in the world to fly from a lunar orbit to the orbit in the vicinity of L2 point \cite{Liu2014}. In our analysis we use the experience of Spektr-RG operation \cite{Pavlinskiy2018}.

Thus, as the analyzed orbital configurations the following were chosen: a halo orbit around the L2 point of the Sun-Earth system with an exit from the ecliptic plane of no more than 400000 km; a circular near-Earth orbit of 40000 km and two high elliptical near-Earth orbits with perigee of 40000 and 100000 km. These orbits were chosen based on the requirements imposed both by the technical constraints of the spacecraft and by the requirements of the scientific program taking into account the experience of previous missions. Configuration of halo orbit around L2 point of the Sun-Earth systems combined with such an exit from the ecliptic plane is based on the experience of Spektr-RG mission and it will provide a better spacecraft visibility to cover about 90\% of the celestial sphere per year with the required periodicity for repeated scientific observations. It will make it possible to observe a larger number of sources comparing to near-Earth orbits and ensure the spacecraft thermal regime. While the near-Earth orbits arise from the experience of space-VLBI missions (HALCA, Astro-G and Radioastron) with the constraint on the minimal safe height of the near-Earth orbits must be not less than 40000 km to ensure the required thermal regime of the spacecraft. The calculated orbits and its parameters are presented Section \ref{section:orbits_calc}.

\section{Methods}\label{section:methods}
Since the requirements of the single-dish mode are clear and can be met in L2 point orbit, in this article we have assessed primarily the capabilities of Millimetron space observatory in VLBI mode for several orbital configurations by analyzing such characteristics as $(u,v)$ coverage, the shape and the relative magnitude of synthesized beam, observational capabilities for different orbital configurations.

General approach that is used in spacecraft orbital calculations is a numerical integration with various force models. Accuracy of integrated orbits for space missions should be at least 1~cm \cite{Papanikolau_2016}. There are plenty of different factors that influence the cumulative forces acting on a spacecraft in Earth orbit. For example, along with the central component of the Earth gravitational field there are nonlinear effects that affect the motion of a satellite in its orbit \cite{Schall_2011}. Also, one should account for the contribution of forces arising from the attraction of other celestial bodies, such as the Moon, the Sun and other. 

\subsection{Orbit Integration}\label{subsection:propagate}
Designing the trajectory of a spacecraft during the mission planning can be highlighted as a separate task. As a rule, the solution of such a problem consists of two stages: 1) the first approximation, 2) the refinement of  obtained approximation using numerical methods. As such methods, various methods of numerical integration of differential equations of motion are used. There are methods of single-step and multi-step integration with constant or variable step. The most widespread of them are the various modifications of the Runge-Kutta method or the Dormand-Prince method that provide the required accuracy with the proper computational speed \cite{Butcher_2008}. Each of theses methods has its own advantages and disadvantages. 

At present, there is a various software that allow, to one extent or another, to solve the problems of designing and predicting the spacecraft trajectories. Among them there are, for example, GMAT\footnote{GMAT software: \url{https://software.nasa.gov/software/GSC-17177-1}}, STK\footnote{STK software: \url{https://www.agi.com/products/stk}}.

However, in a number of tasks, the use of a third-party software products lead to a significant complication of the trajectory design process. In this regard, the use of domestic implementation of the methods of numerical integration is the most practical way. It allows to create more flexible and adaptive software when solving the problem of the trajectory design.

This work presents the results of the software implementation of the Runge-Kutta method up to 7(8)$^{th}$ order with a variable adaptive step with the possibility of flexible adjustment of the force model, which includes:
\begin{enumerate}
 \item Central gravitational field of the Earth;
 \item Non-central effects of the Earth gravitational field (up to the 8$^{th}$ order inclusively);
 \item Gravitational attraction of the solar system bodies: the Sun, the Moon, Mercury, Venus, Mars, Jupiter, Saturn, Uranus, Neptune and Pluto;
\end{enumerate}

The accuracy of the resulting orbits calculated taking into account the above model should be sufficient for the primary analysis and simulations of VLBI observations described in this paper. More accurate force model is supposed to be used in the final calculation of the flight version of the nominal orbit for Millimetron space observatory.

\subsection{Trajectory Design}\label{subsection:orbitdesign}
The considered orbits (elliptical near-Earth orbits, including a circular one and a halo orbit around the Lagrange point L2) are the solution of different problems. In the first case it is a limited two-body problem, in the second it is a limited three-body problem. So the calculation processes are different.

In order to construct elliptical orbits around the Earth an initial vector, set in Keplerian orbital elements, is translated into J2000 coordinate system. Then, it is integrated with a variable step, with an initial step of 1 s. Perturbations are introduced by the force model described above.

To construct an orbit around the L2 point of the Sun-Earth system, the problem of three bodies is considered: the Sun, the Earth, and the spacecraft. This is a non-linear problem, so numerical integration is used to solve it. To construct the orbit, an initial approximation is used, in this case, a linear approximation, then this initial state is integrated forward for the required period. The propagator's capabilities allow for long-term calculations without significant loss of accuracy on the turn.

The success of any space-VLBI mission directly depends on the accuracy of orbit determination. High-precision determination of the state vector of the spacecraft is difficult for the case of a halo orbit around the L2 point of the "Sun-Earth" system, located at a distance of 1.5 million km from the Earth. At such a distance, it is not possible to use the spacecraft laser positioning method due to the large signal losses \cite{SLR}. Therefore, astrometric, radio range and Doppler measurements are used to determine the position and the velocity of the spacecraft. They are the input to differential orbit refinement methods such as Least Squares and Kalman Filter \cite{LSE1,Kal1}. These methods make it possible to compare the measurement model and real observations and reconstruct the true orbit. In addition, inaccuracies in determining the position of the spacecraft can be eliminated by searching for the VLBI fringe at lower frequencies. One way or another, as a result, it is possible to estimate the spacecraft state vector for a certain epoch, and then, using numerical integration, to predict its further orbital motion. In turn, for an accurate forecast, it is necessary to have the most complete force model acting on the space observatory. At the moment, separate publications are being prepared in our laboratory on the topics of prediction and reconstruction of the Millimetron space observatory orbit including the results of development and testing of the corresponding software packages.

It is planned to carry out trajectory measurements for the observatory in X-band range with a maximum error of 20 m in the slant range and 0.5 mm/s in the radial velocity, which will be quite enough to determine the residual delays and successfully obtain correlation for 230 GHz.

\subsection{Fuel Consumption}\label{subsection:fuel}
The flight to L2 point lasts about 100 days. Corrections are made during the flight to reach the nominal orbit. They are appointed for 10, 20 and 40 days. A budget of 100 m/s is allocated for these corrections. Further, to maintain the orbit at the L2 Lagrange point, corrections are carried out approximately every 50 days. The total cost of maintaining corrections is about 15 m/s per year \cite{Nazirov2019}. The offloading of the flywheels was not taken into account.

Two stages for the Millimetron space observatory spacecraft mean it is necessary to understand whether we can change orbits during the lifetime of the project. Two impulses are required to reach a given high elliptical orbit. The first will lead to deorbiting L2 halo-orbit. The second is performed at the perigee of a given near-earth orbit to form an ellipse. To do this, firstly, we estimated what impulse it is necessary to give to the vehicle for deorbiting from L2 halo-orbit into near-earth orbit, as well as fuel costs.

To find the time interval for deorbiting L2 we sampled several points in the orbit over the course of a year, added a small impulse and integrated the orbit forward in time, trying to find where the transfer path would intersect with the given Earth orbit with the same inclination. Then we determined the impulse for the transition to a given orbit around the Earth using coordinate descent algorithm. After that we calculated impulse to form an elliptical orbit by velocity difference.

Impulse costs were recalculated into fuel costs using the following parameters: the initial mass of the spacecraft is 10 tons, the specific impulse of the engine is 2842.29 m/s.

\subsection{Synthesized Beam}\label{subsection:beam}
Synthesized beam reflects the instantaneous response of interferometer for a point-source with 1 Jy flux density. It can be represented as a Fourier transform from the sampling function $S(u, v)$, where the sampling function corresponds to the $(u,v)$ coverage. With the increasing number of independent antennas $N$ the beam shape will be approaching Gaussian shape.

In general, the beam size and shape can be used to some extent to characterize the quality of VLBI interferometer configuration and the quality of image reconstruction. This is a simple, but effective way to preliminary evaluate the VLBI configuration for a given interferometer geometry. 
The beam shape tells not only about the obtainable angular resolution, but also in some measure provides the information of $(u,v)$ coverage quality through the evaluation of the sidelobes effect, which are related to the gaps in the $(u,v)$ coverage leading to the loss of spatial sensitivity in the configured VLBI array \cite{Thompson2017}. In case of space-Earth VLBI the size of the gaps can be substantial \cite{Liu2020}.

\subsection{Source Visibility}\label{subsection:visibility}
The analysis was based on a list of compact extragalactic sources that were proposed as possible candidates for the observations with Millimetron. Most of these sources were observed and had positive detection in Radioastron space-ground VLBI mission \citep{Kardashev2013, Kardashev2015, Kovalev2020}. 
The list included 386 sources. Coordinates were taken from Radio Fundamental Catalog\footnote{Radio Fundamental Catalog: \url{http://astrogeo.org/rfc/}}, J2000.0 epoch.

As ground-based telescopes, we have compiled several sets from all radio telescopes currently available in the world, dividing them into according to the supported frequency bands: 43, 86, 120, 230, 320, 570 GHz. The sets include many telescopes located at around the world, including antennas of largest VLBI networks: EVN, EHT, GMVA, EAVN, VLBA, etc.

The visibility of the sources was calculated over 4 years. Sources were considered observable and visible by a telescope if they had an elevation of more than $7^{\circ}$ above the horizon. A topocentric coordinate system was introduced using the WGS-84 geodetic system. For Millimetron, the following restrictions were imposed on the visibility of sources: the height of the sources from the ecliptic plane should be in the range of $\{-80^{\circ}, 80^{\circ}\}$, while observations were considered possible only in the anti-Solar hemisphere. In addition restrictions were imposed related to the overlapping of the source by the Earth and the Moon. A source was considered unobservable if the difference between the directions to it and to the center of the overlapping body (the Earth or the Moon) was less than 1.1 of the disk radius of this body.   

\section{Results}
\subsection{Orbits and Fuel Estimates}\label{section:orbits_calc}
Using the methods described in Section~\ref{section:methods} we have calculated the following set of orbits for Millimetron space observatory that were considered in Section~\ref{section:orbits}: high elliptical near-Earth orbits with the perigee of 40000 km and 100000 km, circular near-Earth orbits with the semi-major axis of 40000 km optimized for Sgr~A* and M87 imaging correspondingly, L2 point halo orbit. Regarding the circular and high elliptical orbit the minimal height of 40000 km as it was mentioned is dictated by the requirements for thermal regime of the observatory. Flying below this altitude the observatory won't be able to keep the required temperatures. Parameters of calculated near-Earth orbits are shown in Table \ref{table:orbits}.

\begin{table}[ht]
    \begin{center}
    \caption{Parameters of calculated near-Earth orbits for Millimetron space observatory.}
    \label{table:orbits}
    \begin{tabular}{|l|c|c|r|r|r|}
        \hline
        Orbit                     & $a$, (km) & $e$ & $i$, ($^{\circ}$) & $\Omega$, ($^{\circ}$) & $\omega$, ($^{\circ}$)\\
        \hline
        Circular (40000 km)                 & 40000  &  0  &  80 &  70 & -70\\
        \hline
        HEO (Perigee 40000 km)              & 150000 & 0.7 & 100 &   0 & -15\\
        \hline
        HEO (Perigee 100000 km)             & 150000 & 0.3 &  90 &  15 &   0\\
        \hline
    \end{tabular}
    \end{center}
    \raggedright
    \small{$a$ -- semi-major axis, $e$ -- eccentricity, $i$ -- inclination, $\Omega$ --  longitude of the ascending node, $\omega$ -- argument of periapsis.}\\
\end{table}

\begin{figure}[htp]
    \center
    \includegraphics[width=\linewidth]{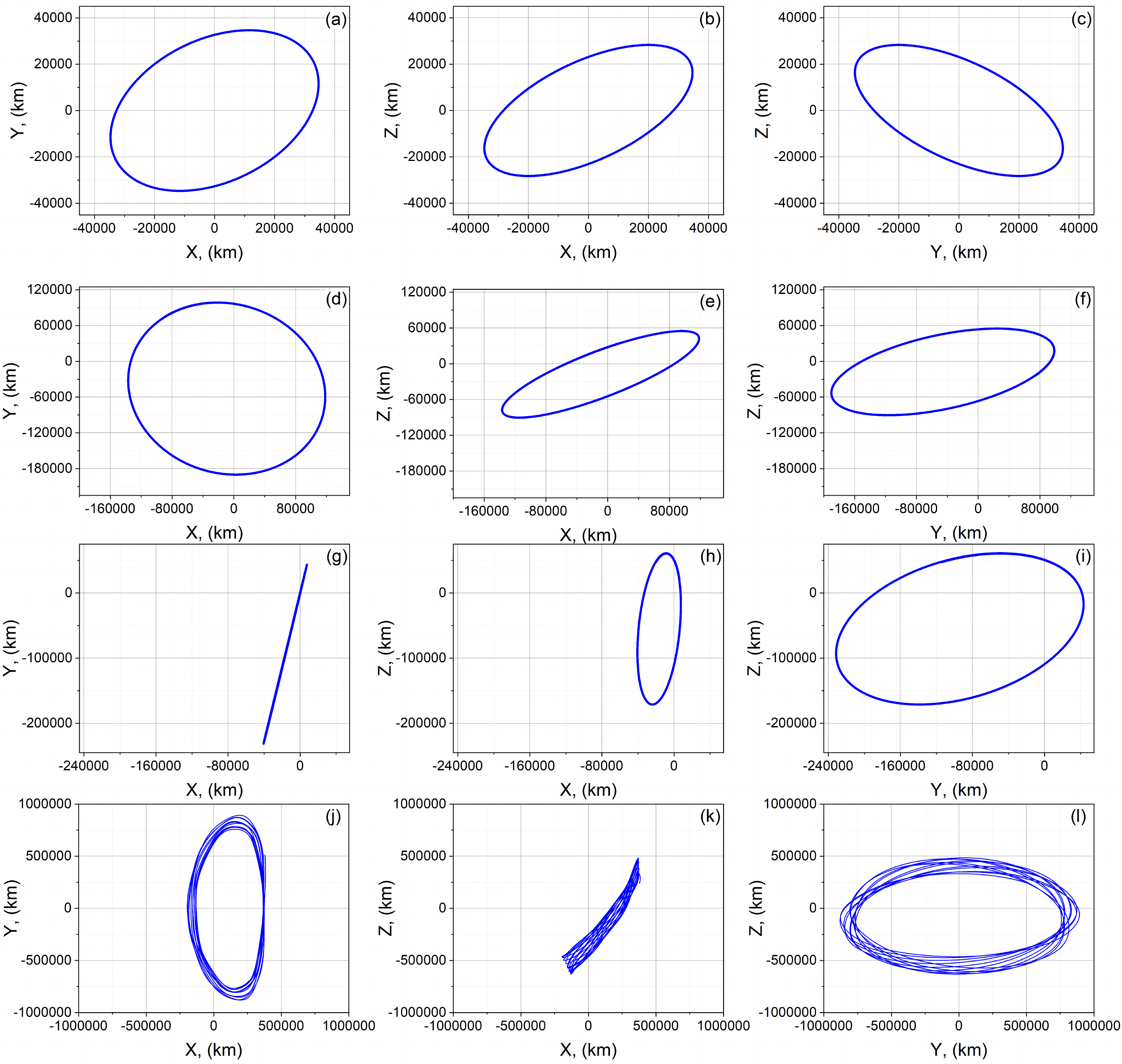}
    \caption{Projections of calculated orbits: circular orbit -- XY (a), XZ (b), YZ (c); high elliptical orbit with perigee 40000 km -- XY (d), XZ (e), YZ (f); high elliptical orbit with perigee 100000 km -- XY (g), XZ (h), YZ (i); orbit around L2 point -- XY (j), XZ (k), YZ (l). Coordinate system is geocentric J2000 for near-Earth orbits and L2-centric for L2 point orbit.}
    \label{fig:fig0}
\end{figure}

The L2 halo orbit period is 1 year, exit from ecliptic plane is 400000 km. This value provides better  spacecraft radio visibility from the tracking stations located on the territory of the Russian Federation.
Fig. \ref{fig:fig0} demonstrates projections of the calculated orbits in geocentric J2000 coordinate system for near-Earth orbits and L2-centric coordinate system for L2 point orbit. The latter coordinate system has it origin in L2 point of the Sun-Earth system, the X-axis is directed towards the Sun, the Z-axis is perpendicular to the ecliptic plane in the direction to the Northern ecliptic pole and the Y-axis complements to a triple.

The next step was to estimate the fuel costs for the calculated orbits. The results of estimating the cost of the flight are presented in Table \ref{table:fuel}. Corrections on the flight in L2 are associated with errors in insertion into the launch orbit, so the budget for the flight will be greater than the budget to return . On the return flight, you may not need any correction at all. The time for maintaining the orbit at L2 point was chosen to be 3 years, since this is the time of active cooling of the main mirror of the spacecraft. Deorbiting in L2 is carried out by a short impulse and does not incur large expenditures of fuel. The magnitude of the impulse depends on the specific orbits. In our case, these are impulses of the order of 1 m/s. Near-Earth orbit formation occurs in perigee of the orbit.

\begin{table}[ht]
    \caption{Impulse and Fuel Estimates.}
    \label{table:fuel}
    \centering
    \begin{tabular}{|l|r|r|r|r|}
        \hline
        \multicolumn{2}{|l|}{Maneuver}        & Impulse, (m/s)         & Fuel, (kg)\\
        \hline
        \multicolumn{2}{|l|}{Correction on flight to the halo-orbit}       & 100  & 358\\
        \hline
        \multicolumn{2}{|l|}{Maintaining the halo-orbit (3 years)}      & 45  & 154\\
        \hline
        \multicolumn{2}{|l|}{Deorbiting the halo-orbit}     & 0-10  & 0-33\\
        \hline
        \multicolumn{2}{|l|}{Correction on flight to the near-earth orbit}  & 0-30  & 0-100\\
        \hline
        \multicolumn{4}{|l|}{Near-Earth orbit formation} \\
        \hline
        \multicolumn{2}{|r|}{Circular (40000 km)} & 1250 & 5160\\
         \hline
         \multicolumn{2}{|r|}{HEO (Perigee 40000 km)} & 265 & 914\\
         \hline
        \multicolumn{2}{|r|}{HEO (Perigee 100000 km)} & 420 & 1490\\
        \hline
    \end{tabular}
\end{table}

As shown in Table \ref{table:fuel} the cost of forming an orbit around the Earth requires a significant amount of fuel. Whereas we have only 800 kg of fuel at our disposal. Therefore, the classical method described in Section \ref{subsection:fuel} is not applicable for us. It's worth to pay attention to gravitational maneuvers around the Moon or to change the parameters of high elliptical orbit around the Earth and to consider invariant manifolds. The first one approach requires detailed analysis due to strict spacecraft orientation constraints while for the second one fuel costs will be smaller if the semi-major axis of HEO is larger or the pericentric distance is smaller.

\subsection{(u,v)-coverage and Synthesized Beam}

For each of the orbits, we calculated the $(u,v)$-coverage and synthesized beam with the ground support of telescopes from Event Horizon Collaboration, considering observing central frequency as 230~GHz and the sensitivities from \cite{EHT2019_P2}. The full list of the telescopes that were used in the beam calculation is shown in the Table \ref{table:tels}. This list includes the EHT antennas that are already available and will be available in the nearest future: Kitt Peak and NOEMA (2021), former CARMA antenna (2023).

All calculations were performed in Astro Space Locator software package for VLBI data reduction \cite{Likhachev2020}. To reconstruct the synthesized beams we used uniform weighing with grid size of 1024$\times$1024 corresponding to the field of view 50$\times$50 $\mu$as. As a kernel function we took an average function.

\begin{figure}[htp]
    \center
    \includegraphics[width=\linewidth]{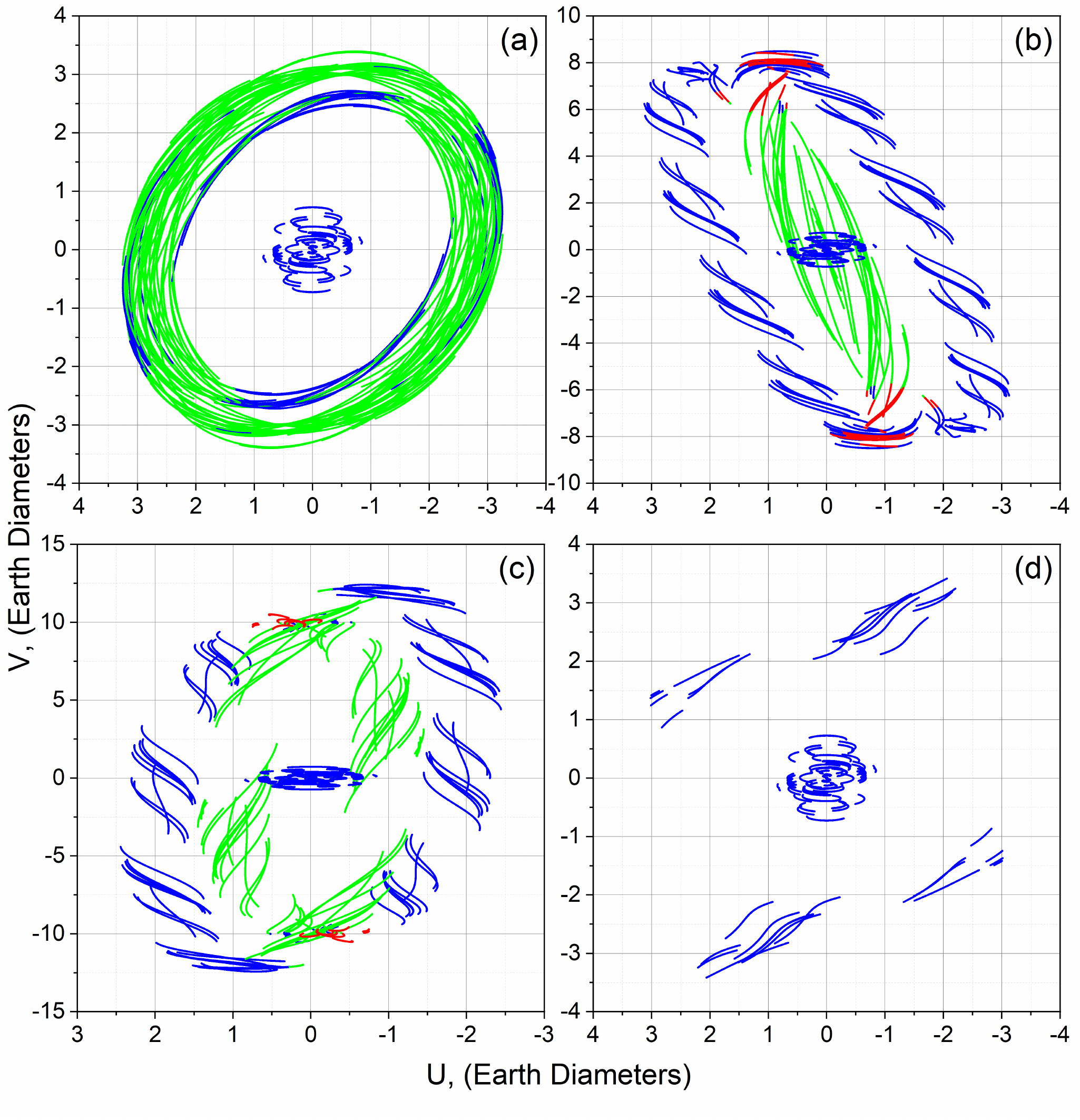}
    \caption{$(u,v)$ coverage for M87 target source: (a) -- circular near-Earth orbit, (b) -- HEO with 40000~km perigee, (c) -- HEO with 100000~km perigee, (d) -- L2 point orbit. Blue dots correspond to observations with no constraints, red dots indicate the Moon illumination, green dots indicate the Earth illumination.}
    \label{fig:fig1}
\end{figure}

\begin{figure}[htp]
    \center
    \includegraphics[width=\linewidth]{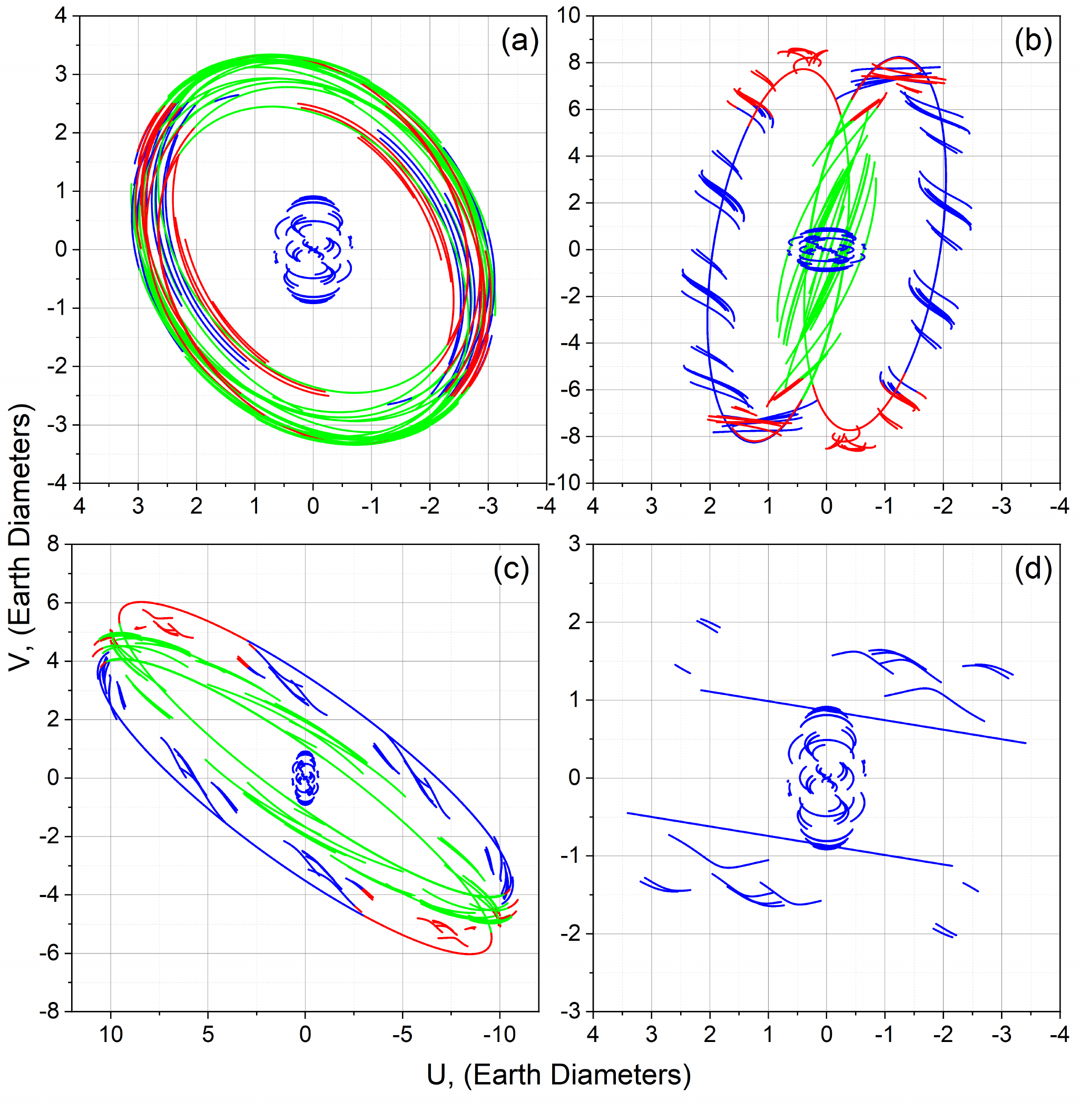}
    \caption{$(u,v)$ coverage for Sgr~A* target source: (a) -- circular near-Earth orbit, (b) -- HEO with 40000~km perigee, (c) -- HEO with 100000~km perigee, (d) -- L2 point orbit. Blue dots correspond to observations with no constraints, red dots indicate the Moon illumination, green dots indicate the Earth illumination.}
    \label{fig:fig2}
\end{figure}

The $(u,v)$ coverage determines the dynamic range of the resulting VLBI image and, in general, its quality and the quality of the synthesized beam. Fig.~\ref{fig:fig1} and \ref{fig:fig2} show $(u,v)$ coverages that were calculated for two priority sources for observations -- M87 and Sgr~A* correspondingly. For near-Earth orbits the accumulated period of time was 4 days for circular orbit, 7 days for HEO orbits (1 orbital period) and 24 hours for L2 point orbit. Blue dots on these figures show the real observations that are not limited by any constraints. As it can be seen the coverage is actually much better for the near-Earth orbits (primarily for circular one, see Fig.~\ref{fig:fig1} (a) and Fig.~\ref{fig:fig2} (a)) than for L2 orbit (Fig.~\ref{fig:fig1} (d) and Fig.~\ref{fig:fig2} (d)).  However, there is a large number of points in near-Earth orbits (especially in HEO, see Fig.~\ref{fig:fig1} and \ref{fig:fig2} (b) and (c)) that will fall under certain constraints which not only will reduce the quality of the coverage, but will probably require additional reorientation of the spacecraft.

\begin{table}[h]
    \caption{The list of telescopes used for beam synthesis.}
    \label{table:tels}
    \centering
    \resizebox{\textwidth}{!}{%
    \begin{tabular}{|c|c|c|c|c|}
    \hline
    \textbf{Telescope}                                                                                           & \textbf{Location}         & \textbf{\begin{tabular}[c]{@{}c@{}}Latitude\\ deg, min, sec\end{tabular}} & \textbf{\begin{tabular}[c]{@{}c@{}}Longitude\\ deg, min, sec\end{tabular}} & \textbf{\begin{tabular}[c]{@{}c@{}}SEFD (for 230 GHz)\\ Jy\end{tabular}} \\ \hline
    Millimetron                                                                                                  & Space                     & -                                                                         & -                                                                          & $\sim$4000                                                               \\ \hline
    \begin{tabular}[c]{@{}c@{}}Atacama Large Millimeter\\ Array (ALMA)\end{tabular}                              & Atacama, Chile            & 23$^{\circ}$01'09''S                                                      & 67$^{\circ}$45'12''W                                                       & 47                                                                     \\ \hline
    \begin{tabular}[c]{@{}c@{}}Atacama Pathfinder\\ Experiment (APEX)\end{tabular}                               & Atacama, Chile            & 23$^{\circ}$00'21''S                                                      & 67$^{\circ}$45'33''W                                                       & 4700                                                                     \\ \hline
    \begin{tabular}[c]{@{}c@{}}Combined Array for Research\\ in Millimeter-wave\\ Astronomy (CARMA)\end{tabular} & California, United States & 37$^{\circ}$16'49''N                                                      & 118$^{\circ}$08'31''W                                                      & 3500                                                                       \\ \hline
    Greenland Telescope (GLT)                                                                                    & Greenland, Danish Realm   & 72$^{\circ}$35'00''N                                                      & 38$^{\circ}$25'00''W                                                       & 5000                                                                     \\ \hline
    \begin{tabular}[c]{@{}c@{}}IRAM 30-m millimeter\\ radio telescope (PV)\end{tabular}                          & Pico Veleta, Spain        & 37$^{\circ}$03'58''N                                                      & 3$^{\circ}$23'34''W                                                        & 1900                                                                     \\ \hline
    \begin{tabular}[c]{@{}c@{}}James Clerk Maxwell\\ Telescope (JCMT)\end{tabular}                               & Hawaii, United States     & 19$^{\circ}$49'22''N                                                      & 155$^{\circ}$28'37''W                                                      & 10500                                                                    \\ \hline
    \begin{tabular}[c]{@{}c@{}}Kitt Peak National\\ Observatory telescope (KP)\end{tabular}                      & Arizona, United States    & 31$^{\circ}$57'30''N                                                      & 111$^{\circ}$35'48''W                                                      & 13000                                                                       \\ \hline
    Large Millimeter Telescope (LMT)                                                                             & Mexico                    & 18$^{\circ}$59'09''N                                                      & 97$^{\circ}$18'53''W                                                       & 4500                                                                       \\ \hline
    Plateau de Bure (PdB)                                                                                        & Alps, France              & 44$^{\circ}$38'02''N                                                      & 5$^{\circ}$54'29''E                                                        & 1600                                                                       \\ \hline
    Submillimeter Array (SMA)                                                                                    & Hawaii, United States     & 19$^{\circ}$49'27''N                                                      & 155$^{\circ}$28'41''W                                                      & 6200                                                                       \\ \hline
    Submillimeter Telescope (SMT)                                                                                & Arizona, United States    & 32$^{\circ}$42'06''N                                                      & 109$^{\circ}$53'28''W                                                      & 17100                                                                       \\ \hline
    South Pole Telescope (SPT)                                                                                   & Antarctica                & 90$^{\circ}$00'00''S                                                      & 00$^{\circ}$00'00''E                                                       & 19300                                                                       \\ \hline
\end{tabular}%
}
\end{table}

\begin{figure}[htp]
    \center
    \includegraphics[width=\linewidth]{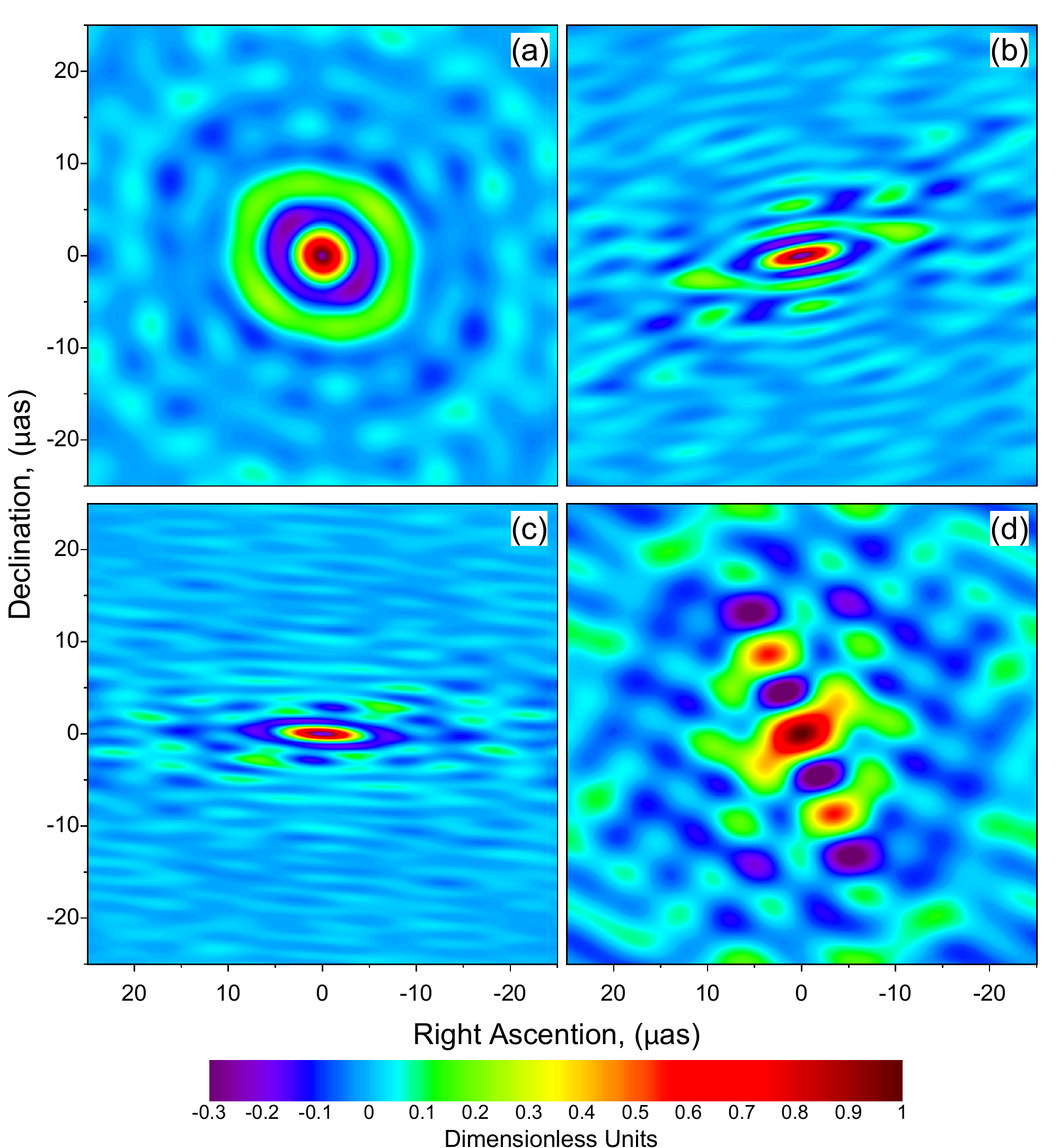}
    \caption{Synthesized beams for M87 target source: (a) -- circular near-Earth orbit, (b) -- HEO with 40000 km perigee, (c) -- HEO with 100000 km perigee, (d) -- L2 point orbit. The intensity is normalized to the peak value.}
    \label{fig:fig3}
\end{figure}

\begin{figure}[htp]
    \center
    \includegraphics[width=\linewidth]{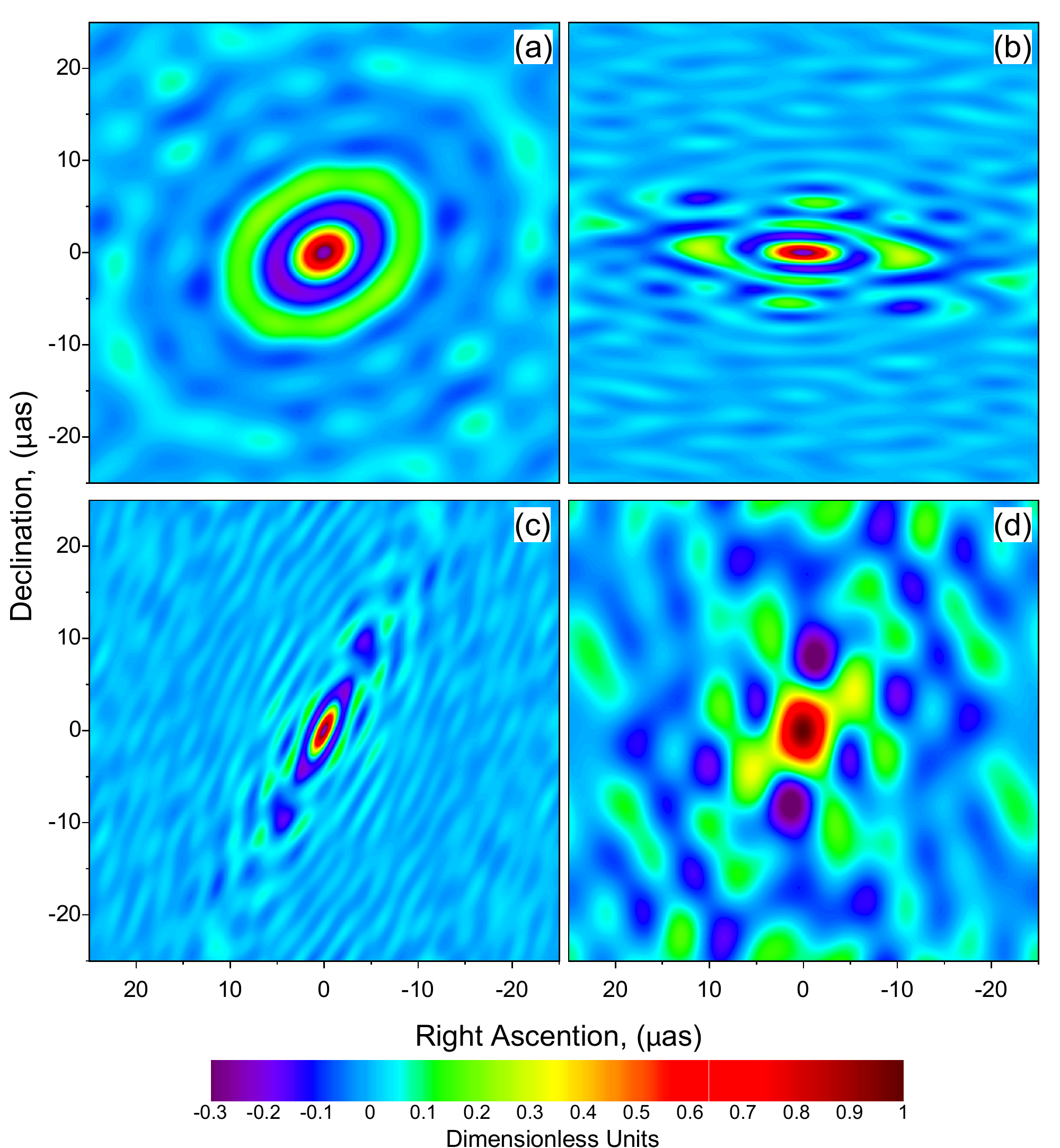}
    \caption{Synthesized beams for Sgr~A* target source: (a) -- circular near-Earth orbit, (b) -- HEO with 40000 km perigee, (c) -- HEO with 100000 km perigee, (d) -- L2 point orbit. The intensity is normalized to the peak value.}
    \label{fig:fig4}
\end{figure}

\begin{table}[ht]
    \caption{Beam parameters.}
    \label{table:beam}
    \centering
    \begin{tabular}{|l|r|r|}
        \hline
        Orbit                                   & Beam size, ($\mu$as)\\
        \hline
        Circular (40000 km) M87                 &  4.4 $\times$ 4.0 \\
        \hline
        Circular (40000 km) Sgr~A*              &  5.0 $\times$ 3.8 \\
        \hline
        HEO M87 (Perigee 40000 km)              &  6.0 $\times$ 1.6 \\
        \hline
        HEO M87 (Perigee 100000 km)             &  6.2 $\times$ 1.2 \\
        \hline
        HEO Sgr~A* (Perigee 40000 km)           &  5.7 $\times$ 1.6 \\
        \hline
        HEO Sgr~A* (Perigee 100000 km)          &  4.3 $\times$ 1.2 \\
        \hline
        L2 Orbit M87                            &  6.5 $\times$ 4.3 \\
        \hline
        L2 Orbit Sgr~A*                         &  6.9 $\times$ 4.9 \\
        \hline
    \end{tabular}
\end{table}

Image recovery procedure implies sequential subtracting of the beam from the maxima of the dirty map (for more details see \cite{Thompson2017}, Section 11.1 and \cite{Hogbom1974}). Based on this principle, it can be concluded that beam configuration for the corresponding $(u,v)$-coverage defines the resulting quality of image recovery. For example, in case of large sidelobes in a beam false components could arise, creating artifacts in the resulting image. Asymmetrical elongated beam make it difficult to recover images. Based on this we can conclude that the best beam will be a symmetrical (circular) beam with dimensions at half maximum about 5 $\mu$as in the case of mapping sources such as SMBH, for example, Sgr~A* or M87).

Fig. \ref{fig:fig3} and \ref{fig:fig4} show the obtained beams for different orbits for M87 and Sgr~A* sources correspondingly. Obviously, the best radial symmetry is achieved with a circular orbit (Fig. \ref{fig:fig3}, \ref{fig:fig4} (a)). In case of a high elliptical orbit (Fig. \ref{fig:fig3}, \ref{fig:fig4} (b) and (c)), it is radially elongated in one of the directions. For halo orbit around the L2 point (Fig. \ref{fig:fig3}, \ref{fig:fig4} (d)), the beam has a number of interesting properties. It has a radial symmetry comparing to HEO orbit, albeit not as much as for a circular one. At the same time, its dimensions allow to perform imaging with a resolution of about 5 $\mu$as (see Table \ref{table:beam}). Achievable angular resolution of each orbit configuration presented in Table \ref{tab:angular}.

\begin{table*}[htp]
\centering
\caption{The best and the worst angular resolution achievable for considered orbits.}
\label{tab:angular}
\begin{tabular}{|l|c|c|c|c|c|}
\hline
\multirow{2}{*}{Orbit} & \multicolumn{5}{c|}{Angular Resolution (best - worst), $\mu$as}  \\ 
\cline{2-6} 
                       & 43 GHz      & 90 GHz      & 240 GHz     & 330 GHz    & 570 GHz   \\ 
\hline
Circular         & 36 -- 133   & 18  -- 56   & 5.9 -- 20   & 4.8 -- 15  & 2.2  -- 7.0     \\ 
\hline
HEO              & 4.5 -- 4500 & 2.1 -- 2145 & 0.8 -- 800  & 0.6 -- 585 & 0.27 -- 280     \\
\hline
L2               & 0.8 -- 550  & 0.4 -- 265  & 0.14 -- 100 & 0.1 -- 72  & 0.05 -- 35      \\
\hline
\end{tabular}
\end{table*}

Maximum baseline projection for ground array of the EHT at 230~GHz reached the values of about 25 $\mu$as \cite{EHT_2019P1}, while space VLBI with the Radioastron yield the resolution of 8 $\mu$as at 1.3 cm \cite{Kovalev2020}. The resolution from the estimated beams for considered orbital configurations demonstrates the increase in angular resolution for imaging comparing to the previous observations reaching $\approx$ 5 $\mu$as at 230~GHz for all orbits. Speaking about the best achievable angular resolution L2 orbit will provide the highest angular resolution less than 1 $\mu$as. However this is true for the short survey observations and not for the full imaging. 

\subsubsection{Snapshot Mode}
The snapshot $(u,v)$ coverage is critical to imaging science for intra-day variable sources like Sgr~A*. The point is that a detailed study of the dynamics of the close vicinity of black holes, the structure of which can be rapidly changing (for example, Sgr~A* $\tau\sim221$ s, $\approx10 GM/c^{3}$, see \cite{Moscibrodzka2014,Moscibrodzka2016}), requires a qualitative instantaneous (snapshot) $(u,v)$ coverage which should accumulate in a short time comparable to the source variability. Fig. \ref{fig:fig6} shows the resulting snapshot $(u,v)$ coverages for four orbital configurations: circular, two highly-elliptical orbits and L2 point orbit.

\begin{figure}[ht]
    \center
    \includegraphics[width=\linewidth]{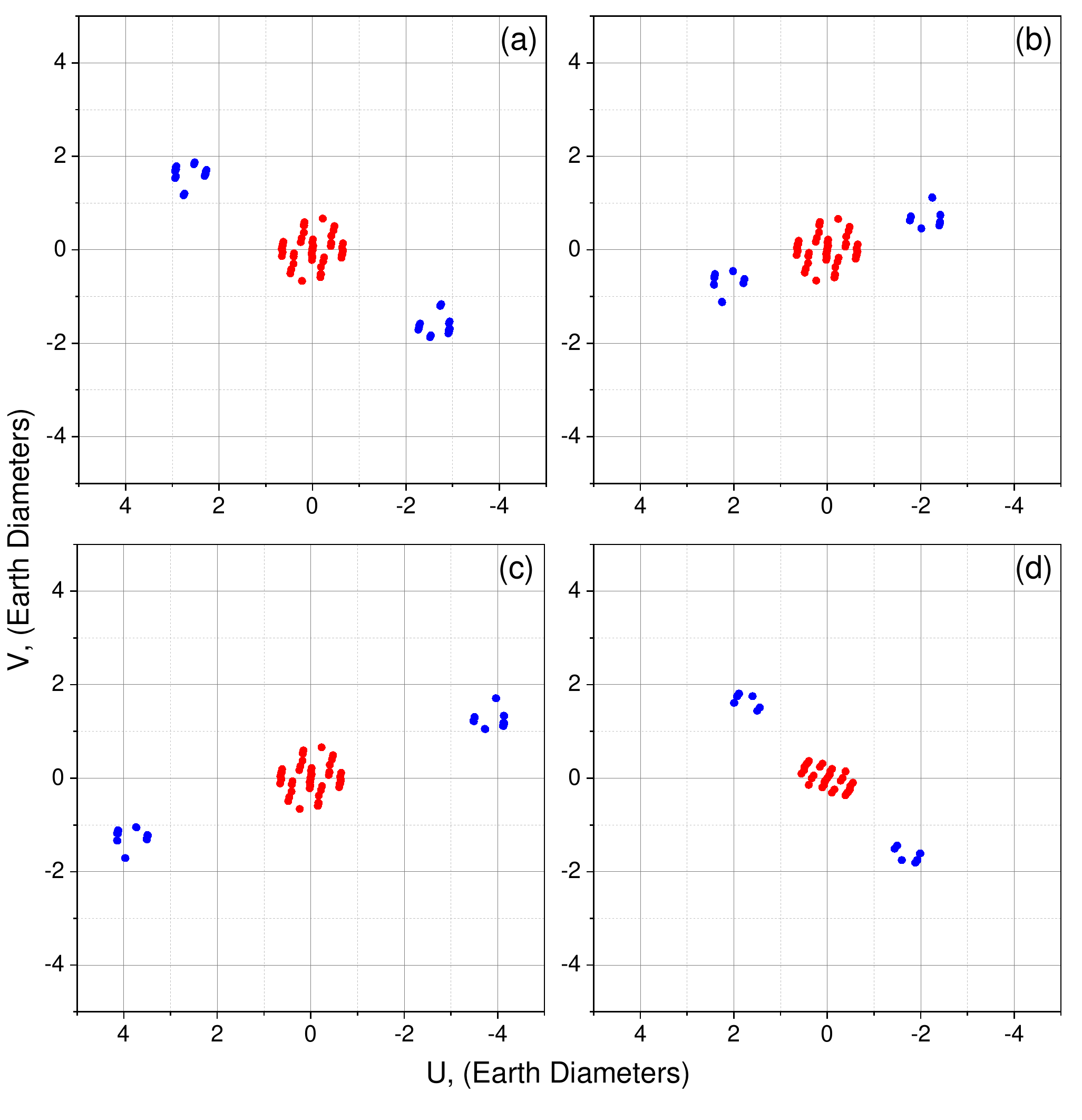}
    \caption{Snapshot $(u,v)$ coverage for (a) -- circular near-Earth orbit, (b) -- high elliptical near-Earth orbit with 40000 km perigee, (c) -- high elliptical near-Earth orbit with 100000 km perigee, (d) -- orbit around L2 point of the Sun-Earth system.}
    \label{fig:fig5}
\end{figure}

The calculated snapshots had a duration of $\sim 4$ minutes. However, we were unable to restore the synthesized beam as all of the $(u,v)$ coverages are too sparse. Previously, in \cite{Andrianov2020} a possibility to perform space-VLBI dynamical observations of Sgr~A* was demonstrated, but for different HEO configuration with a low perigee of $\sim 10000$ km. The thermal regime requirements do not allow implementation of an orbit with such a low perigee.
Orbital configurations considered in this work are not capable of dynamic observations of such variable sources as Sgr~A* and suggest only averaged observations.

\subsubsection{Multi-band Observations}
\begin{figure}[ht]
    \center
    \includegraphics[width=\linewidth]{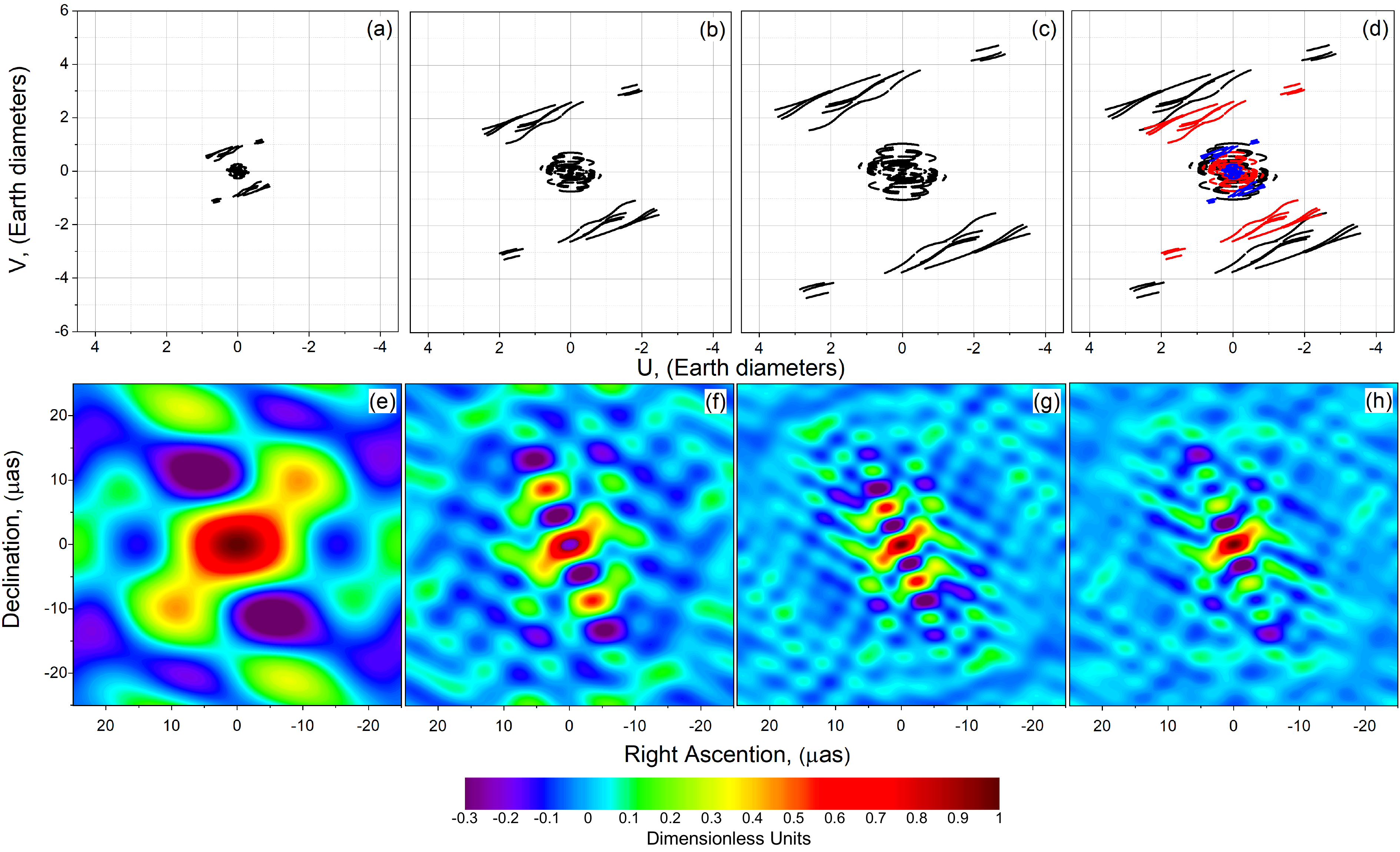}
    \caption{$(u,v)$ coverage and synthesized beam for 86 (a and e), 230 GHz (b and f), 345 GHz (c and g) and for multi-band 86+230+345 GHz (d and h) observations. Plot (d): blue dots correspond to 86 GHz observations, red dots correspond to 230 GHz observations and black dots correspond to 345 GHz observations.}
    \label{fig:fig6}
\end{figure}

Space-ground VLBI differs from the ground-based observations with its rather sparse $(u,v)$ coverage. However, one way to improve both the coverage and resolution is to use multi-frequency observation \cite{Blackburn2019,Johnson2019p}. Millimetron space observatory will be capable to observe at several frequencies simultaneously. Therefore we have estimated the $(u,v)$ coverage for 86+230+345 GHz observations with Millimetron and EHT ground telescopes for the case of L2 point orbit. The list of considered frequencies corresponds to the project for the further development of the EHT (new generation EHT or ngEHT) where multi-frequency ground-based VLBI are also assumed.
Below Fig.~\ref{fig:fig6} (a), (b), (c) and (d) show the $(u,v)$ coverages for 86 GHz, 230 GHz, 345 GHz and multi-frequency (86+230+345 GHz) observations and Fig.~\ref{fig:fig6} (e), (f), (g) and (h) show the corresponding synthesized beams.

Adding up two more frequencies improves the situation with the side-lobes (less expressed side-lobes for multi-frequency observations) and the angular resolution: 14.5 $\times$ 10 $\mu$as, 6.5 $\times$ 4.3 $\mu$as, 4.45 $\times$ 2.8 $\mu$as beams for 86 GHz, 230 GHz, 340 GHz correspondingly and 4.7 $\times$ 2.9 $\mu$as for multi-frequency. Moreover, as it can be seen from Fig.\ref{fig:fig6} (h), the 86 GHz space-ground baselines will overlap the 230 GHz ground $(u,v)$-coverage.

In addition to the improvement of $(u,v)$ coverage and resolution, multi-frequency observations can help to increase the coherent integration time using phase transfer methods which is valuable for the ground part of millimeter wavelength range VLBI observations \cite{Rioja2017,Zhao2018}.

\subsection{Source Visibility}
Both for L2 and near-Earth orbits only 7 sources from the initial list are not visible: 0506-612, 0637-752, 1642+690, 1656+602, 1807+698, 1823+689, 1842+681, 1849+670, 1933+655. Thus, 379 out of 386 considered sources are available for space-ground observations with Millimetron.

\begin{figure}[ht]
    \center
    \includegraphics[width=\linewidth]{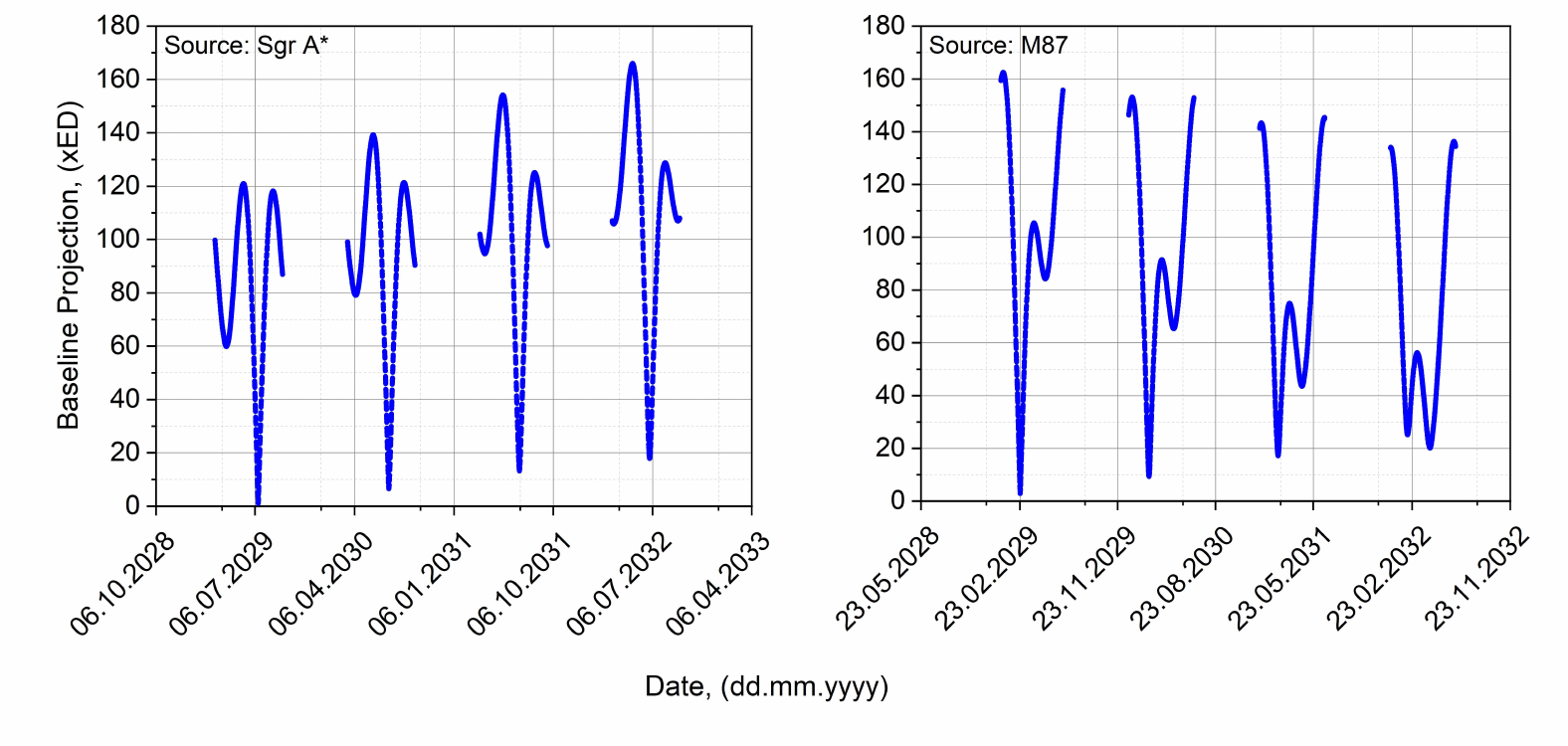}
    \caption{Baseline projection in time for Sgr A* (left) and M87 (right) for L2 point orbit. Time span -- 4 years. Gaps correspond to the case when the sources are not observable due to the Sun constraints.}
    \label{fig:fig7}
\end{figure}

Fig. \ref{fig:fig7} shows the annual visibility windows for Sgr~A* and M87 sources for 4 years of mission operation. Blank regions correspond to the case when the sources are not available for observations due to the Sun constraints. Thus, VLBI observations at short baseline projections are possible once per year.

\begin{figure}[ht]
    \center
    \includegraphics[width=\linewidth]{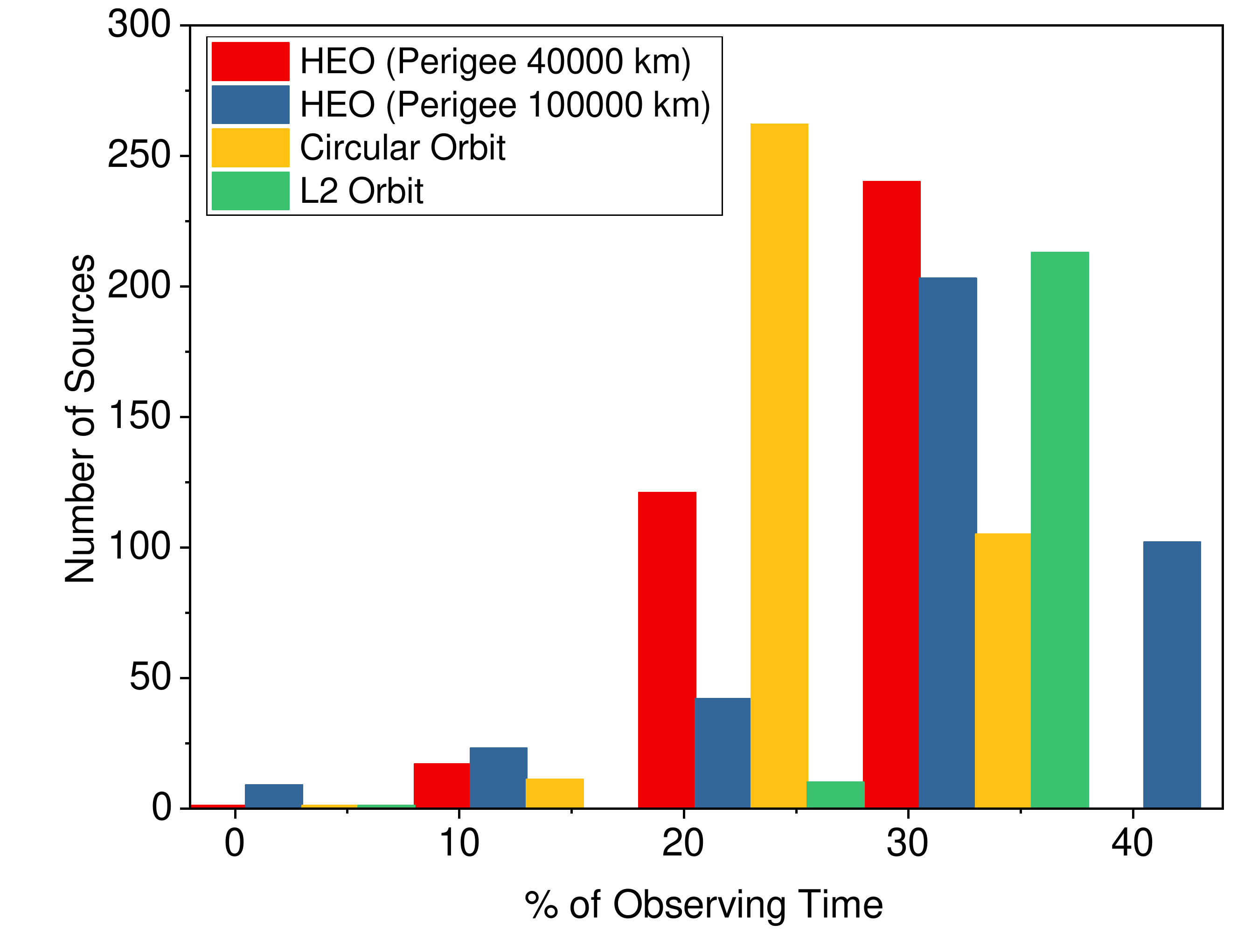}
    \caption{Number of sources vs. percentage of available observing time for them for 4 years of mission operation.}
    \label{fig:fig8}
\end{figure}

Fig.~\ref{fig:fig8} shows the number of sources depending on the percentage of observable time during 4 years of mission operation. L2 orbit provide larger number of sources that are observable in 40\% of 4 years of mission operation time, while for near-Earth orbits (HEO and circular) there are additional constraints of the Sun, the Moon and the Earth that reduce the total observing time to 20-30\%.

Although the considered orbits were optimized primarily for observations of two target sources -- Sgr~A* and M87, there is also the possibility of observing and imaging the sources that are close to the orbital plane of the spacecraft. These could be for example: M84, 3С454 or OJ287.

\section{Discussion and Conclusions}\label{section:conclusion}
We have calculated several orbital configurations: circular near-Earth orbit with radius of 40000 km, two highly elliptical orbits with 40000 and 100000 km perigee and L2 orbit with an exit from the ecliptic plane of 400000 km.

Due to the requirements for the thermal regime of the antenna and onboard scientific payload of the single-dish mode, the orbit around L2 point of Sun-Earth is considered as the main orbit of the Millimetron observatory. All other calculated orbits were considered as the possible second stage.

Each orbital configuration then was analyzed from a technical and scientific points of view. This analysis consisted of several steps: evaluation of synthesized beam, quantitative estimation of the number of observable sources, calculations for impulse and fuel costs for orbital corrections and transfer to/from Sun-Earth L2 point.

From the point of view of VLBI imaging observations, namely the quality of $(u,v)$ coverage, the best orbital configuration is circular orbit. However, the results of fuel costs calculations demonstrate that it is impossible to organize a combined "L2 point+circular near-Earth orbit", while circular orbit itself doesn't meet the requirements for the single-dish observations due to the constraints of the Earth, the Sun and the Moon and their regular exposure of the observatory.

High elliptical near-Earth orbits have a potential to perform the transfer to/from the L2 point of Sun-Earth system using the gravitational maneuvers. For our fuel estimations we considered a direct transfer to/from L2 point and in this case HEO orbits are still above the limits of the amount of available fuel. Thermal regime of the observatory applies constraint on the perigee -- it must be not lower than $\sim$ 40000 km leading to drop in the $(u,v)$ coverage quality for VLBI imaging and making such dynamic observations that are mentioned in \cite{Andrianov2020} impossible. Here we did not provide an analysis of the possibility of transferring between orbits using gravity assist maneuvers, since this is a separate study performed out of the scope of this work.

From the point of view of the single-dish observations, the best possible orbital configuration for Millimetron observatory will be orbit around L2 point of Sun-Earth system. At he same time we have demonstrated that L2 point orbit can provide shorter baselines (less than 5 Earth diameters) which means that there is a possibility to perform VLBI imaging once per year for several given target sources. Such limit arise from the constraints applied by the Sun. Synthesized beam of L2 point orbit demonstrates a lower quality comparing to the near-Earth orbits due to more sparse $(u,v)$ coverage which however can be improved by multi-band observations.
Moreover, estimated source visibility at L2 point orbit is 30\% better comparing to the near-Earth orbits that have additional constrains from the Sun, the Moon and the Earth.

Thus, the calculated orbit around L2 point is now suggested as a the new nominal orbit. We believe that optimizing such L2 point orbit is more practical approach than developing a two stage mission program with transfer between the orbits. 

Regarding the imaging capabilities of different orbital configurations, a more detailed analysis can be found in \cite{Andrianov2020, Likhachev2021} where various models of sources are considered and the sensitivity of the interferometer is analyzed.

It should also be borne in mind that the calculated orbits belong to their families of orbits and may be changed and optimized depending on the final spacecraft launch date and the final scientific program and requirements.

\section*{References}
\bibliography{biblio}
\end{document}